\documentclass[conference]{IEEEtran}
\usepackage{amsmath,amssymb,amsfonts}
\usepackage{algorithmic}
\usepackage{graphicx}
\usepackage{textcomp}
\usepackage{xcolor}
\def\BibTeX{{\rm B\kern-.05em{\sc i\kern-.025em b}\kern-.08em
    T\kern-.1667em\lower.7ex\hbox{E}\kern-.125emX}}

\usepackage{url}
\usepackage[hidelinks]{hyperref}
\usepackage{cleveref}
\usepackage{physics}

\usepackage[numbers]{natbib} % Used in combination with \bibliographystyle{IEEEtranN} below

\begin{document}

\title{Distributed Quantum Computing:\\ Applications and Challenges}

% Following 5 lines necessary for proper aligning author names
\makeatletter
\newcommand{\linebreakand}{%
  \end{@IEEEauthorhalign}
  \hfill\mbox{}\par
  \mbox{}\hfill\begin{@IEEEauthorhalign}
}
\makeatother

\author{\IEEEauthorblockN{Juan C. Boschero, Niels M. P. Neumann, Ward van der Schoot, Thom Sijpesteijn, Robert Wezeman}
\IEEEauthorblockA{\textit{Dept. Applied Cryptography \& Quantum Algorithms} \\
\textit{The Netherlands Organisation for Applied scientific Research (TNO)}\\
The Hague, The Netherlands \\
}}

\maketitle

\begin{abstract}
Quantum computing is presently undergoing rapid development to achieve a significant speedup promised in certain applications.
Nonetheless, scaling quantum computers remains a formidable engineering challenge, prompting exploration of alternative methods to achieve the promised quantum advantage.
An example is given by the concept of distributed quantum computing, which aims to scale quantum computers through the linking of different individual quantum computers. 
Additionally, distributed quantum computing opens the way to new applications on the longer term. 
This study seeks to give an overview of this technology on an application-level, considering both use cases and implementation considerations.
% In this overview, specific use cases in which distributed quantum computations provides an advantage over local classical and quantum computations are listed. 
% Moreover, the paper seeks to illuminate the influence of distributed quantum computing on implementing quantum algorithms by providing an extensive overview of the crucial implementation considerations.
In this way, this work aims to push forward the field of distributed quantum computing, aiming for real-world distributed quantum systems in the near future.
\end{abstract}

% Keywords 

\section{Introduction}
Computing power has grown exponentially since the early 1960s following the predictions of Moore's law, which asserts that the number of transistors on a computer chip doubles roughly every two years~\cite{Moore}. 
Unfortunately, it is starting to become apparent that this exponential growth is reaching its limit.
The already minuscule sizes of transistors limit the unbounded increase of the number of transistors on chips. 
Luckily, a new form of computing, called quantum computing, has arisen recently, which promises to continue this exponential growth.

Quantum computing is a form of computing based on quantum mechanics, a theory of physics relevant at the smallest scales. 
Quantum mechanics describes phenomena such as superposition and entanglement, which do not arise classically. 
Hence, quantum computers can perform fundamentally different operations than the ones performed by classical devices. 
With clever design of these operations, quantum computing makes it possible to reach up to exponential speedups in various fields, such as machine learning, chemistry and optimisation~\cite{Uses_QC}.

Quantum computing is currently under rapid development with an apparent exponential scaling in the number of qubits, the basic building blocks.
This scaling is similar to what Moore's law predicted for classical computers~\cite{IBM}, and what classical computers are currently following.
Many applications require thousands of qubits for quantum computing to become of practical relevance, especially when accounting for decoherence effects.
While modern quantum devices can achieve up to 1000 qubits~\cite{IBM}, the quality of these qubits remains low. Numerous applications require qubits and operations with extremely low error rates.
Currently available quantum devices, often referred to as noisy intermediate-scale quantum (NISQ) devices~\cite{preskill2018}, are noisy and suffer from loss of information.

Because of this, practical large-scale algorithms are still years away, even with the current scaling of qubits and error rates.
Instead of scaling the resources on a single device, multiple quantum devices can be combined to attain increased computational power with the given quantum devices. 
This idea is called \textit{distributed quantum computing} (DQC), and works similarly to the classical distributed settings often used in high-performance computing (HPC).
Depending on the noise rates of quantum communication links, DQC offers a possible faster path to scaling quantum hardware~\cite{IBM:classical_communication:2022}. 
Additionally, DQC opens the path to new applications where different parties can collaborate and thereby solve more complex problems than when running algorithms themselves. 

The first ideas of DQC are already applied to scale current quantum hardware~\cite{Davarzani:2022}.
An example are the Heron chips by IBM, which use the chiplet architecture~\cite{laracuente2023modelingshortrangemicrowavenetworks}, in which communication between different chips is used to scale the overall computational power~\cite{IBM,IBM:classical_communication:2022}.
Furthermore, one upcoming technique is given by circuit cutting, in which larger circuits are cut up into smaller circuits to run on smaller quantum hardware~\cite{Mitarai_2021,Mitarai_2021a, Tang_2021}. While this is not the same as DQC, it is the same idea of using techniques to solve larger problems on a quantum computer than one would expect based on the resources alone~\cite{Piveteau:2023,Rennela:2023}.

There are different ways of achieving DQC. 
In the broadest definition, DQC can be seen as solving a problem using multiple quantum chips or devices.
The devices can be in close proximity, situated on separate chips within the same computer, or they can be located kilometres apart.
This definition also includes settings where different quantum devices independently solve separate parts of a problem, and where the solutions to the individual parts are combined to give the answer to the full problem. 
If the outcome of the individual parts is classical, this distributed settings requires only classical communication between the devices.
In this work, the focus lies on the case where the different quantum devices cooperate in a quantum way using quantum communication and where interaction between the devices is necessary before extracting the individual answers. 
Because of that, DQC is more specifically defined as the execution of a quantum circuit distributed over different parties, in which at least one gate is distributed over different devices, and applied through quantum communication.

Quantum communication entails the sharing of quantum information between different parties.
While it is a key ingredient for DQC, it also plays a vital role in other applications, such as Quantum Key Distribution~\cite{Bennett2014,QKD} and the quantum internet~\cite{QuantumInternetCaccia, Kimble_2008, doi:10.1126/science.aam9288}. 
These two examples specifically use quantum communication to transmit information from one party to another.
In DQC, however, quantum communication is a prerequisite for performing joint computations. 

Generally speaking, there are two advantages of using DQC over quantum computing on a single device. 
The first advantage has already been mentioned above: by allowing cooperation between different quantum devices, the total amount of computing resources increases, resulting in a more powerful computational unit. 
One could enhance computational power in the near future by leveraging the diverse strengths of various quantum technologies through the combination of devices based on different technologies, thus advancing DQC capabilities.
The second advantage is that it allows for collaborative computing. 
If the different quantum devices are controlled by different parties, each party could input their own data on their respective quantum device.
DQC would then allow these parties to perform operations on their joint data, thus allowing collaborative insights among the parties.
In the long term, DQC thus opens the way to completely new applications. 

Until recently, most of the theoretical research in the field of DQC has been focused on distributing specific algorithms. In experimental studies, the emphasis has been on very fundamental aspects such as distributing gates across systems at distances ranging from a few meters~\cite{60m_optical_fiber} to several kilometers~\cite{liu2023distributed}.
Recently, a couple of works have started to give an overview of the field of DQC, with works shown in~\cite{caleffi2022distributed} and~\cite{barral2024reviewdistributedquantumcomputing}, being most notable.
In the first survey, the authors provide an overview of DQC by looking at four different perspectives: algorithms, networking, compiling and simulation. 
The work focuses mostly on the low-level challenges associated with these different aspects of DQC. 
The second work gives a full overview of the state of DQC from a rather technical perspective. It is a great overview for anyone trying to understand the workings of DQC, touching upon hardware, architecture, compilation and algorithm design.

In contrast, the present work aims to give an overview of the application-level aspects of DQC.
Specifically, we focus on end-users and applications, detailing what DQC implies in practice and what the associated challenges and advantages are.
Typically, a close interaction with classical high performance computers is required when running quantum algorithms. Emphasizing the forward-looking nature of this work is crucial, given that the existing DQC infrastructure is still in its early experimental stages. 
Furthermore, as this problem (and its solution) extends to distributed settings, we will not explicitly discuss this integration. 

This work is structured as follows. In Section~\ref{section:types}, we introduce two notions of DQC and explain their differences. 
We then discuss different use cases for both types of DQC in Section~\ref{section:usecases}.
In Section~\ref{section:aspects}, we detail some (technical) considerations around the development and practice of DQC.
In particular, we discuss how quantum algorithms, quantum communication and quantum hardware are relevant to a distributed framework. 
Lastly, we conclude our overview with a discussion and outlook on the future of DQC. 

\section{Types of DQC}
\label{section:types}
% Reasons for doing this
As mentioned in the introduction, there are two main reasons to opt for distributed operations.
These reasons extend beyond the field of DQC.
The first reason is that your own computational resources do not suffice for solving the problem. 
The second reason is that you do not have all data available for solving the problem. 
Both reasons translate to different applications and infrastructures. 
Below we name and explain both types in the quantum setting, after which we discuss potential applications for both in the next chapter. 

% For larger problem sizes
We call the first type \textit{resource DQC}. 
Here, the available quantum resources are insufficient or unsuitable for running the algorithm and we turn to other parties or devices to collectively achieve the required resources.
This type closely relates to a client-server model in classical distributed settings~\cite{Benatallah:2004}. 
The used resources can even originate from different servers.
From the perspective of the client, no difference should exist between one server and multiple linked servers. 
The insufficient resources can differ in type and extend beyond just an insufficient number of qubits.
For certain algorithms, it could be beneficial to distribute (part of) the computation so that one can take advantage of hardware properties of the other device.
For example, reducing the circuit depth by distributing the computation to a device with a larger gate set or a better qubit connectivity.  

% For collaborating
We call the second type \textit{data DQC}.
Here, the data is distributed over multiple parties and quantum computing and quantum communication can link the data to perform computations on all data collaboratively.
This type of distributed computing closely relates to classical multi-party computation~\cite{Canetti:1996}, with the added benefit that quantum computations can directly be performed on the data~\cite{Neumann2022}. 

Encoding data onto quantum computers is not a straightforward process. It is an active area of study, typically involving the initialisation or preparation of quantum states at the start of a circuit to align with classical tensor data \cite{prepare_quantum_state}, or utilising quantum feature maps to encode classical inputs as quantum data \cite{lloyd2020quantum}. This embedded data would subsequently be communicated through entangled qubits to other devices, as detailed in Section \ref{section:quantum_com_architecture}.

The two types of DQC may also appear combined and in a hierarchical setting. 
For instance, in a data DQC setting, one party may implement resource DQC to perform their local computations. 
Table~\ref{tab:types_DQD} contains these two types of DQC, as well as regular quantum computing, and lists whether the resources and the data are local or distributed for each entry. 
Note that a setting with only local quantum resources and distributed data falls outside our scope for DQC (i.e., with at least one distributed quantum gate) and is thus omitted. 
\begin{table}[!t]
\centering
    \caption{Nomenclature for different types of distributed QC}
\begin{tabular}{l|l|l}
\textbf{Paradigm name} & \textbf{Resources} & \textbf{Data} \\ \hline 
Regular QC & local & local \\
Resource DQC & distributed & local \\
Data DQC & distributed & distributed 
\end{tabular}
    \label{tab:types_DQD}
\end{table}

%An example of where resource DQC comes into play is quantum cloud computing~\cite{NSS:2023}. 
%In this setting, users have only remote access to a quantum computer on which to perform computations. 
%All local resources, including the data, are classical. 

% Ways of doing this
% Cloud-based quantum computing
%% Full quantum versus classical-quantum hybrid
%% Small quantum-large quantum hybrid

\section{Different Use Cases}
\label{section:usecases}
This section describes four potential applications of DQC.
As practically any quantum algorithm can benefit from resource DQC, we stress that the list is incomplete. 
It should, however, be mentioned that certain applications are more naturally suitable for DQC than others.
The next chapter gives some indication as to what makes an algorithm suitable for DQC.
The four presented use cases are examples of suitable algorithms and provide insights in the potential of DQC. 

\textbf{Quantum Machine Learning:} 
% Training without explicitly sharing data (work with Thom, with Robert)
%% QML + Arithmetic
%% Maybe split this one up
Machine learning (ML) is a leading contemporary research domain with applications in diverse fields, such as image and video recognition, natural language processing, recommendation systems, and fraud detection~\cite{Sarker2021}.
In its simplest form, machine learning focuses on training models with a certain amount of training data, such that the resulting model can be applied to unseen data to predict relevant properties related to this data.
These ML tools have transcended academia and are now accessible to the general public, exemplified by instances like DALL·E2~\cite{dall-e2}, which generates images from text descriptions, and ChatGPT~\cite{chatgpt}, a conversational artificial intelligence model capable of engaging in human-like interactions. 
The downside of these models is that their continual upkeep and refinement requires processing large data sets, resulting in the use of significant computational resources.
The number of floating-point operations needed to train state-of-the-art models has been shown to be growing exponentially~\cite{Narayanan2021}, which can quickly become a problem because of the unrealistic long training times.

Quantum Machine Learning (QML) is the field that tries to combine quantum computing with machine learning. 
A potential advantage of QML is speeding up computation, i.e., the quantum algorithm that solves the specific ML problem runs faster than a classical alternative algorithm.
A possible speedup is however not the only dimension that needs to be considered for assessing true meaningful advantages of QML. 
For an elaborate discussion we refer to chapter 9 of~\cite{schuld2021machine}.
% performing the same and handling data more efficiently~\cite{schuld2021machine}.

Both forms of DQC defined in Section~\ref{section:types} can be beneficial for QML. 
Resource DQC can facilitate the implementation of larger QML models that surpass the qubit capacity of an individual quantum device. 
Thus, this type of DQC can accommodate the execution of more complex models demanding increased qubit connectivity or adherence to a specialised gate-set.  
Data DQC, on the other hand, allows different parties to supply input to an QML algorithm collaboratively. 
This can be useful when different parties own different parts of the data that are needed to train a model.
In the context of classification, this could for example be the case when one party possesses particular types of distorted images of an object, while another party knows the actual type of object due to having access to other resources such as sound. 
Combining their data, these two parties are able to train a ML model that is able to classify the distorted images.

In the NISQ era, so-called \emph{variational quantum algorithms} (VQAs) are promising methods of quantum machine learning. 
VQAs are hybrid quantum-classical algorithms, where a parameterised circuit is run on a quantum computer and the parameter optimisation occurs in a classical outer loop.
Two well-known VQAs are the highly structured QAOA~\cite{farhi2014quantum} and the Variational Quantum Eigensolver (VQE)~\cite{dewolf2023quantum}. 
VQAs are expected to be especially relevant in the NISQ era, as many believe the parameters can be optimised to accommodate for the noise present. 
In addition, because of their simple form, variational algorithms require only few qubits and small depth compared to more traditional fault-tolerant quantum algorithms~\cite{Cerezo_2021}. 
Because of this, many if not most of the current quantum machine learning approaches are based on variational algorithms.

Recently, distributing quantum computations in the context of VQAs has seen its first research papers.
Two examples are the resource DQC version of VQE by~\cite{DiAdamo_2021} and~\cite{khait2023variational}. 
Many other research in distributed versions of VQAs tends to distribute the classical part of the computation (e.g.,~\cite{distributed_learning_scheme_vqas,large_scale_qaoa,Niu2023}), but this falls outside the scope of this present paper.

\textbf{Secure computations:} % Performing large computations securely on a remote server
%% Maybe be more explicit here
% Efficient distribution of complex gates, e.g. Molmer-Sorensen gate with all-to-all connectivity
%% Maybe be more explicit here
%% Storing states
%% Combining strength of different devices
% Others?
\textit{Multi-party computation} refers to a field of research with the goal to allow two or more parties to perform analyses or computations on their joint data without revealing their own data. Security is a vital requirement for many applications in multi-party computation.
This concerns the information exchanged between different parties and that the exchanged data cannot be exploited to disclose any unintended information about either party.  

Quantum computers offer inherent security of quantum states, as quantum states are destroyed upon measurement. 
This naturally makes quantum technologies suitable for applications in security.
Most quantum algorithms will be hybrid, i.e., combining classical and quantum routines, such as variational algorithms. 
The inherent security does not transfer to the classical routines, hence resulting in potential security issues.
However, these security issues also arise the equivalent classical algorithms.
The use of quantum only increases the security of the algorithm, specifically for the information stored in quantum states.

DQC adds extra security to the implementation of quantum computers. 
Firstly, in the context of resource DQC, it removes the need for sending input and algorithm data to the quantum computer provider.
Currently, most quantum computers can only be accessed through cloud providers. 
To use cloud-based quantum computers, users have to send all instructions to the quantum computer provider and that way essentially share all information about their input and their algorithm.
This removes all potential security offered by quantum devices.
By using resource DQC, input and algorithm data only need to be inputted to a local quantum device, after which DQC allows the use of the cloud-based quantum computer and its required extra resources.
No data is shared with the cloud provider because of the security obtained through quantum states~\cite{jozsa2005introduction}.

Secondly, in the context of data DQC, multi-party computation are possible without explicit data sharing between the different parties. 
In such settings, each party inputs their data into their local quantum device, after which a distributed computation is performed. 
Then, no single party directly learns data that is inputted by other parties. In addition, the algorithm can be modified so that no party learns what algorithm was actually run and which problem was actually solved, both while executing the algorithm, as well as when measuring the qubits and processing the measurement outcomes. 
Note however that each party does learn some part of the algorithm and quantum state, so no hard guarantees can be given on the security resulting from this way of distributing quantum computations. 
The classical communication required between quantum devices, however, does not leak the quantum state information of the remote qubits. 

% Depending on the availability of a small local quantum device, the situation may simplify.
% If a local quantum device is available which can be linked to the cloud-based quantum computer, quantum teleportation and gate-teleportation routines can be used to implement the gates, where the information shared only depends on the local measurement outcomes unknown to the quantum host.

%On the other hand, if no local quantum device is available, opting for post-quantum cryptographical solutions might yield a solution. 
%The quantum computer provider is asked to prepare a specific quantum state and measure part of the qubits. 
%Based on the measurement outcomes, the user knows what quantum state is available, whereas the post-quantum cryptographic protocols prevents the quantum computer provider from learning anything about the outcome of the algorithm~\cite{Cojocaru2021}. 
%Even though the applied operations are known to the quantum computer provider, the initial state is not and hence the measurement outcome lacks context for correct interpretation. 

%Note that in general, no hard guarantees on the security resulting from this way of distributing quantum computations can be given. Each provider does learn part of the algorithm and part of the initial state. The communication required with other quantum devices does furthermore not leak information on the remote part of the quantum state.

\textbf{Breaking cryptography:}
Cryptography is a field of research that enables many processes in ICT, such as secure communication, data protection and authenticated messaging. 
The algorithms to protect these processes depend on certain mathematical problems which are believed to be hard to solve using current classical computers. 
Because of this, cryptography has managed to protect vital ICT infrastructures for over forty years.

Quantum computers will have a significant impact, as they prove valuable in breaking cryptographic protocols.
Grover's algorithm~\cite{Grover:1996} weakens symmetric cryptography using a quadratically faster unstructured search algorithm, whereas Shor's algorithm~\cite{Shor_1997} breaks most asymmetric cryptography by exploiting the structure used in the protocol to achieve an exponential speed-up. 

Both algorithms could benefit from a resource DQC implementation. Grover's algorithm is not naturally suited for DQC implementations, as its traditional implementation requires many qubits communicating with one another. However, an exact version of the algorithm exists where many function evaluations are performed in parallel~\cite{Zhou2023}. This version requires fewer qubits communicating with one another, making the algorithm more suitable for distributed settings.
%In this version, every increase in width decreases the depth with the same factor, which follows from a lower bound on the query complexity of unstructured search~\cite{Boyer1998}. 
% Given many parallel instances, running Grover's algorithm in a resource DQC setting could help in breaking cryptography using unstructured search. 
% The advantage of distributing Grover's algorithm likely lies in scaling the number of qubits more easily.

The situation is quite different for Shor's algorithm.
The phase estimation part of the algorithm consists of multiple controlled phase gates that all commute. 
In addition, only certain qubits need to communicate with one another in the algorithm.
These two characteristics make it that Shor's algorithm is naturally suited for distributed settings.
% A single extra device can therefore reduce the depth by a factor two. 
% The depth and width of Shor's algorithm scale similarly to Grover's algorithm. 
% For Shor's this depth is however linear in the input and distributing the algorithm is therefore easier. 
The work of~\cite{Yimsiriwattana:Shor:2004} shows a possible implementation of a distributed Shor's algorithm and discusses the potential speedup. 

Data DQC seems to be of no relevance for both algorithms, as neither of them requires input data distributed over or owned by different parties.

\textbf{Quantum interferometry:}
Interferometers are instruments that utilise the interference of superimposed waves for extracting information.
These instruments find applications in various fields~\cite{HARIHARAN2007}. 

Current interferometers convert the information from the incoming wave packets in classical data, which proved optimal for applications requiring small amounts of interferometers measuring a source with a high signal to noise ratio~\cite{HARIHARAN2007}. 
However, classical computations with interferometric data tend to encounter challenges, particularly in fields where the signal-to-noise ratio is low and multiple interferometers are required, like astronomy~\cite{Boffin2016}.
Quantum computers can solve issues in such scenarios as the quantum information of a wave package, or photon, can be directly transcribed onto a qubit resulting in less resolution noise.
This gives rise to new quantum interferometers.

If one were to have access to different interferometers, the same signal could be measured from different locations and in more detail. 
A central interferometer could then combine the light incoming from the separate interferometers, after which the central interferometer extracts the required information.
By having access to data from multiple interferometers, this extracted information would be more accurate.
However, classical methods require the incoming light to be brought together physically, which is susceptible to noise.

Quantumly, the case is different, leading to a data DQC use case.
Given multiple quantum interferometers, the quantum information measured by a single quantum interferometer could directly be translated onto its local quantum device.
Different quantum computers located at the different interferometers can then run a DQC algorithm.
The result of this algorithm yields the required information, without having to physically bring the different signals together.
This shows that data DQC can improve the performance of quantum interferometers.

It should be noted that different to the aforementioned use cases, the advantage of DQC mainly comes from the quantum conversion of information rather than computation. 
Still, computation is required to process the signals, for which a quantum computer is naturally suitable.
Because of this, this use case is still considered as a DQC use case, be it with a different focus.

\section{Technical considerations}
\label{section:aspects}
DQC promises to bring many advantages, yet it is far from an out-of-the-box solution.
Quantum computers themselves already have their limitations and will likely only bring (great) benefits to specific fields. 
Even within these fields, DQC is unlikely to speed up all quantum algorithms. 
When applying DQC, there are different technical considerations to take into account that 
% There are different technical considerations to take into account when using DQC, some of which even pose limitations on the use of DQC. 
influence whether the use case benefits from DQC, and if so, how it should be used specifically.
This section discusses the technical considerations that currently seem most important. 
The considerations are grouped top-down in three categories: algorithms, quantum communication architecture and quantum computing hardware.

\subsection{Algorithms}
Many practical aspects come into play when developing (distributed) quantum algorithms.
This imposes a limit to the sort of quantum algorithms that benefit from running on distributed systems.

First, algorithms that contain many gates between different pairs of qubits are inherently unsuitable for DQC implementations.
Noise in the communication links dominate imperfections in quantum devices and hence form the limiting factor in implementations of DQC.
An efficient distributed quantum algorithm should therefore minimize the amount of communication between different devices. 
Effectively, most operations should be performed on the same device, which is possible if the quxxbits can be split up into multiple groups, so that most of the operations only happen within each of the separate qubit groups.
Algorithms that require many different pairs of qubits to interact, such as in quantum volume circuits~\cite{QuantumVolume:2019}, therefore only marginally benefit from distributing quantum computations.
Specifically for data DQC, this means that algorithms in which the data stored at different quantum computers needs to be communicated between different parties a significant amount of times, are unsuitable for DQC.
On the other hand, algorithms suitable for standard circuit cutting techniques might also be suitable for distributed settings, as the overhead due to the circuit cutting translates roughly to the communication complexity in distributed algorithms. 
One must note that some protocols retain a high success probability in a distributed setting, even with noisy communication links~\cite{RNSW:2023}.

Second, algorithms that contain many blocks of commuting operations are naturally suitable for DQC implementations.
This is due to the fact that commuting operations can be applied in parallel. 
So in DQC settings, every individual quantum computer can run one block of commuting operations in parallel. 
To apply these commuting operations in parallel, there needs to be a way to link the different operations.
This is achieved in~\cite{HS2005} using a non-local quantum fan-out gate, a gate that naturally extends the non-local CNOT gate, introduced in~\cite{Eisert2000}, to multiple targets~\cite{Yimsiriwattana2004}. 
As such, this reduces the depth of some algorithms significantly, as subsequent operations are replaced by parallel ones. 
A prime example is the phase estimation algorithm, which allows us to parallellize Shor's factoring algorithm~\cite{Shor_1997} to constant depth~\cite{HS2005}.

\subsection{Quantum Communication Architecture} \label{section:quantum_com_architecture}
To turn single-device quantum computing into DQC, connections between different quantum devices are required. 
Different considerations have to be considered when setting up such connections. 
These considerations concern either the quantum network topology as a whole, or the coupling of specific quantum devices. 
The following two subsections will detail the considerations concerning both.
\begin{figure*}[h] 
\centering
    \includegraphics[width=1\textwidth]{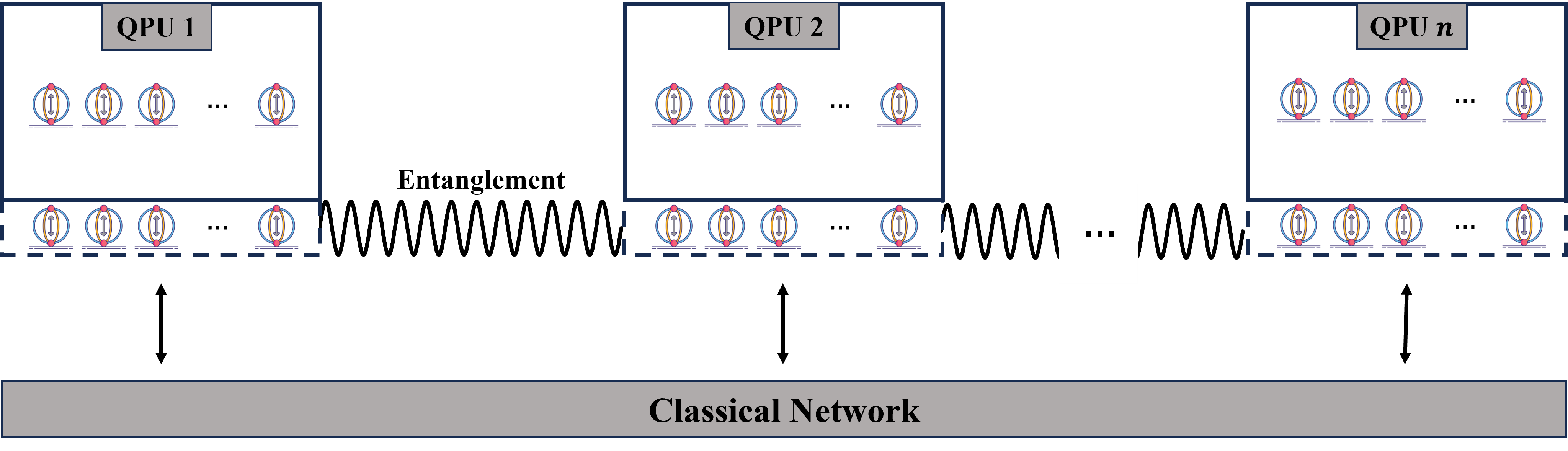}
    \caption{Illustration of a fundamental quantum computing network topology featuring \textit{n} quantum computers. Each square symbolizes an abstracted quantum computing processing unit (QPU), equipped with an unspecified number of computational qubits within the solid square, and an additional qubit within each dashed box, denoting the communication qubit. These communication qubits facilitate the establishment of entanglement between individual QPUs. Additionally, the QPUs are interconnected via a classical network, essential for executing distributed algorithms.}
    \label{fig:dqc}
\end{figure*}
\subsubsection{Quantum Network Topology}
%% http://staff.ustc.edu.cn/~kpxue/paper/IEEENetwork-zhaoyingwang-202209.pdf
% In coupling quantum devices with one another, various considerations exist to take into account.
% The considerations almost exclusively revolve around the network of quantum devices. 
In DQC, each quantum device uses two types of qubits:
\textit{Computation qubits} used to perform the quantum algorithm; 
and, \textit{communication qubits} used to communicate with other quantum devices.
With the communication qubits, we can set up entanglement between different devices. Therefore, a quantum computer must forfeit at least one qubit to establish communication with another quantum computer, rendering it unavailable for directly computing the algorithm.Moreover, the quantum devices necessitate a pre-established classical network among them to execute distributed algorithms. This is illustrated in Figure \ref{fig:dqc}, which depicts a fundamental quantum network topology between quantum computers.

Which qubit is used for which task is an important decision, as the total number of qubits is limited per device and the coherence times of the qubits is still limited. 

For each pair of entangled communication qubits between two devices, the two devices can perform a joint non-local 2-qubit gate. 
After the 2-qubit gate has been performed, the entanglement must be set up again before it can be used for a new non-local operation.
With more communication qubits, users can perform more non-local operations before having to regenerate entanglement between the communication qubits. 
As a down side, the computational power of a device decreases, as less computation qubits are available. 

For each quantum device, some qubits are more suited for communication, whereas others are better for computation. 
Which qubit is best for which task depends on for instance the number of local (SWAP) operations required and the fidelity of the qubits when connecting them to a remote quantum network. 
The optimal choice can even depend on time~\cite{asynchronous_network}. 
In addition, during a quantum algorithm, a qubit can change between being a communication or computation qubit, to accommodate which role fits best for which qubit at any moment. 

A closely related question is which quantum devices are directly coupled in the setting with more than two quantum devices,.
If two quantum devices have a shared entangled pair of qubits, this pair can be used to perform non-local quantum operations between them directly.
If two devices do not have a shared entangled pair of qubits, non-local quantum operations can only be performed through a chain of entangled qubits on other quantum devices or via entanglement swapping operations.
The non-local operation will then destroy all pairs of entangled qubits. 
A direct quantum link between the device thus saves entanglement generation efforts. 
On the other side,  every quantum connection between two different quantum devices requires a fresh communication qubit.
The choice for which quantum links exist between which devices is also something that can be optimised during the execution of an algorithm. 

Both considerations need to be taken into account when setting up a distributed quantum system. 
For example, a designer can reduce the total number of communication qubits by not directly connecting two quantum devices when only a few non-local operations need to be implemented between them. 
On the other hand, a designer can choose to set up many quantum connections between quantum devices that have to perform many non-local operations between them.
Again, the moment in time, the quantum algorithm, and the individual quantum computer characteristics influence which devices are directly connected and how many qubits will be used for each connection.
\subsubsection{Coupling of quantum devices}
Shared entangled pairs of qubits help couple different quantum devices. 
As different physical phenomena (see next section) can help produce qubits, different physical connections between the different types of qubits are necessary. 

In general, we distinguish between direct and indirect coupling. 
Direct coupling happens when the two quantum devices directly interact which each other, resulting in the required entanglement. 
Direct coupling mainly works for resource DQC when the devices are physically close. 
In data DQC, the different quantum systems most likely are physically separated. 
In this case, indirect coupling can be used.
Indirect coupling uss an intermediate system to set up the entanglement.
First, local entanglement between the intermediate system and one of the two devices is set up. 
The qubit of the intermediate system is then transported to the other quantum device, and entangled with a qubit of the other device.
This protocol establishes entanglement between the two quantum devices.

Despite the lower dissipation and decoherence demonstrated by direct coupling~\cite{Zhang_2011}, indirect coupling currently serves as the primary approach for quantum communication~\cite{Xiang_2013}. 
For the purpose of quantum information transfer, current attention is directed towards optomechanical resonators because they can link the vibrations of a system (phonons) to particles of light (photons).
These photons can subsequently be transmitted through optical fibres~\cite{Hafezi_optical} or free space. 
Quantum information can accurately be transported across 143km in experimental settings~\cite{Herbst_2015}.
In addition, the teleportation of a CNOT gate between logical qubits has been realised~\cite{Chou_2018}.
Superconducting cables can transport qubits by indirectly coupling onto co-planar wave guides connected to the superconducting quantum computers as demonstrated in~\cite{Zhong_2021}, where a GHZ was correctly transported between state between two superconducting quantum computers.

\subsection{Quantum Computing Hardware}
The quantum hardware naturally significantly impacts the implementation of DQC, and different aspects have to be taken into account.
Currently, a wide variety of quantum devices are in development.
These devices differ significantly in for instance used qubit technology, qubit topology, number of qubits and many more. 
In this section, we discuss which of these aspects are relevant for DQC.

With the rise in quantum information science, a wide variety of quantum architectures emerged. 
As of today, the systems that have undergone experimental controllable and coherent quantum manipulation include trapped ions~\cite{Noel_2022}, atoms~\cite{Saffman_2016}, spin systems~\cite{Struck2016} and superconducting circuits~\cite{annurev_soa,orlando_flux}. 
Moreover, among the various quantum architectures currently under development are postulated options such as magnonic qubits~\cite{Lachance_Quirion_2019} and skyrmion qubits~\cite{Psaroudaki_2021}. These quantum systems are built with the specifications set by~\cite{DiVincenzo_2000} in mind, which states that a competitive quantum computer must strive to achieve:
\begin{enumerate}
    \item A scalable physical system with well characterised qubits;
    \item The ability to initialise the qubit states to a simple state, i.e., $\ket{000\dots}$;
    \item Decoherence times (much) longer than gate operation times;
    \item A ``universal" set of quantum gates;
    \item A qubit-specific measurement capability;
    \item The ability to interconvert stationary and flying qubits;
    \item The ability to properly transmit flying qubits from one location to another. 
\end{enumerate}
The first five specifications form the basic requirements of a local quantum computer, while specification 6 and 7 detail the needs for proper communication between quantum systems. 
Different quantum systems excel at different requirements.
As an example, superconducting qubits, and especially transmon qubits, show a considerable advantage for requirements 1 and 2.
Ion trap qubits show considerable advantages for requirement 3.
Currently, no single standardised system exists that consistently outperforms others in \textit{all} aspects.

Note that specifications 6 and 7 are of special relevance to DQC compared to local quantum computations. Although specifications 6 and 7 are heavily quantum communication related, it is important to note that trapped ion systems and photons excel in this field. Transmitting photonic~\cite{Lago_Rivera_2023} and trapped ion~\cite{Palani_2023} qubits has been achieved at long distances using free space or fibre optic cables. 

Although the specifications in~\cite{DiVincenzo_2000} are important for basic functionalities, they do not necessarily translate to practical applicability. To measure how well a quantum device can solve partical problems, quantum metrics are required. Various quantum metrics exist, each with their own focus and use case~\cite{levels_quantum_metrics:2023}.

Both the above specifications and quantum metrics are relevant for the design of DQC algorithms. 
In classical computing, deploying an algorithm across multiple nodes is often unfavourable due to startup costs, security concerns, and the added complexities in overheads. 
However, due to the diverse architectures offered in quantum computing, distributed quantum computing often possesses a larger pay off when compared to its classical counterpart as the advantages of the different architectures can be leveraged. 
Considering the strengths and weaknesses of the different quantum devices, the quantum algorithm can be instantiated such that the strength of each of the devices is utilised, while avoiding each of the weaknesses. 
For example, pairing a computer architecture possessing long coherence times with an architecture with fast operation times allows one system to store information while the other executes the operations of an algorithm.
The mapping of algorithmic qubits to physical qubits should thus take these aspects into account.
It should also be noted that some quantum metrics allow the benchmarking of distributed quantum systems, which could show useful insights when deciding which system to use for which part.

\section{Discussion and outlook}
\label{section:discussion}
In this work, we presented an overview of distributed quantum computing (DQC) and its use cases.
We first identified two types of DQC: 
\textit{Resource} DQC, where the local quantum computation resources are insufficient; 
and, \textit{Data} DQC, where the relevant information is distributed over multiple parties and quantum computers are used to allow computations on the shared data. 
Combinations between the two also exist, as well as hierarchical settings. 

In Section~\ref{section:usecases}, we discussed the benefits that DQC may possess in the areas of quantum machine learning, secure computing, breaking encryption, and quantum interferometry. 
A clear view of the precise value that DQC will bring is still unclear, as it depends on the nature of the hardware used in the future. 

Section~\ref{section:aspects} discussed different technical considerations around quantum computing in a distributed setting.
We discussed the effects of DQC on quantum algorithms, noting that communication will be a bottleneck when distributing algorithms.
The number of devices over which the algorithm is distributed will affect the way performance and the losses due to the communication overhead between the devices, furthermore, more devices give more degrees of freedom to optimally implement the algorithm. 
In the context of quantum communication architecture, we discuss the difference between computation and communication qubits, as well as the consequences of directly and indirectly coupling quantum devices. 
With direct coupling, a quantum link between two devices exists, whereas with indirect coupling, entanglement between two devices has to be set up via one or more other (intermediate) devices. 
We ended the section by considering quantum hardware, noting that DQC is currently only achievable using photon or trapped ion systems as these systems easily allow the transition of in-transit quantum states to quantum states useful for computation.

In the near future, resource DQC offers a solution to scaling up quantum computational power, by linking together multiple smaller quantum devices.
Furthermore, data DQC opens the path to new applications, where multiple parties can collaboratively solve a problem with their combined data. 
Especially data DQC requires quantum devices of higher quality than currently available. 
Currently, the main challenges lie in improving the quantum hardware and improving (the capabilities of) the quantum network. 
The first implementations on actual distributed hardware will however require dedicated implementations and low-level optimizations. 

Distributed quantum computing does introduce an additional overhead, both in the number of required qubits, as well as in the communication between different devices. 
Yet, DQC also offers a new path towards scaling up quantum hardware capabilities and it opens the way to novel applications of quantum computers. 
With this work, we presented a first overview of DQC, incorporating technical considerations as well as algorithmic considerations and use-cases. 
In this way, we hope that this overview forms a unified language for research on DQC, and gives first glimpses into its potential impact and aspects to take into account.
This will hopefully accelerate the field of DQC, paving the way towards proper implementation of DQC in the short and long term.

% Use below to support \citeauthor
\bibliographystyle{IEEEtranN} % In contrast to IEEEtran (ie: without -N), IEEEtranN support citeauthor if combined with natbib. But then, we need to fix the size, so we put footnotesize:
{\footnotesize \bibliography{main}}

% Generated by IEEEtranN.bst, version: 1.14 (2015/08/26)
\begin{thebibliography}{69}
\providecommand{\natexlab}[1]{#1}
\providecommand{\url}[1]{#1}
\csname url@samestyle\endcsname
\providecommand{\newblock}{\relax}
\providecommand{\bibinfo}[2]{#2}
\providecommand{\BIBentrySTDinterwordspacing}{\spaceskip=0pt\relax}
\providecommand{\BIBentryALTinterwordstretchfactor}{4}
\providecommand{\BIBentryALTinterwordspacing}{\spaceskip=\fontdimen2\font plus
\BIBentryALTinterwordstretchfactor\fontdimen3\font minus
  \fontdimen4\font\relax}
\providecommand{\BIBforeignlanguage}[2]{{%
\expandafter\ifx\csname l@#1\endcsname\relax
\typeout{** WARNING: IEEEtranN.bst: No hyphenation pattern has been}%
\typeout{** loaded for the language `#1'. Using the pattern for}%
\typeout{** the default language instead.}%
\else
\language=\csname l@#1\endcsname
\fi
#2}}
\providecommand{\BIBdecl}{\relax}
\BIBdecl

\bibitem[Moore(1965)]{Moore}
G.~E. Moore, ``Cramming more components onto integrated circuits,''
  \emph{Electronics}, vol.~38, no.~8, pp. 114--117, 4 1965.

\bibitem[Hassija et~al.(2020)Hassija, Chamola, Goyal, Kanhere, and
  Guizani]{Uses_QC}
V.~Hassija, V.~Chamola, A.~Goyal, S.~S. Kanhere, and N.~Guizani, ``Forthcoming
  applications of quantum computing: peeking into the future,'' \emph{IET
  Quantum Communication}, vol.~1, no.~2, pp. 35--41, 2020.

\bibitem[Gambetta(2023)]{IBM}
J.~Gambetta, ``{IBM Quantum Computing: Roadmap},'' 12 2023.

\bibitem[Preskill(2018)]{preskill2018}
J.~Preskill, ``Quantum {C}omputing in the {NISQ} era and beyond,''
  \emph{{Quantum}}, vol.~2, p.~79, Aug. 2018.

\bibitem[IBM(2022)]{IBM:classical_communication:2022}
IBM. (2022, 05) {Expanding the IBM Quantum roadmap to anticipate the future of
  quantum-centric supercomputing}.

\bibitem[Davarzani et~al.(2022)Davarzani, Zomorodi, and
  Houshmand]{Davarzani:2022}
\BIBentryALTinterwordspacing
Z.~Davarzani, M.~Zomorodi, and M.~Houshmand, ``A hierarchical approach for
  building distributed quantum systems,'' \emph{Scientific Reports}, vol.~12,
  no.~1, 9 2022. [Online]. Available:
  \url{http://dx.doi.org/10.1038/s41598-022-18989-w}
\BIBentrySTDinterwordspacing

\bibitem[LaRacuente et~al.(2023)LaRacuente, Smith, Imany, Silverman, and
  Chong]{laracuente2023modelingshortrangemicrowavenetworks}
\BIBentryALTinterwordspacing
N.~LaRacuente, K.~N. Smith, P.~Imany, K.~L. Silverman, and F.~T. Chong,
  ``Modeling short-range microwave networks to scale superconducting quantum
  computation,'' 2023. [Online]. Available:
  \url{https://arxiv.org/abs/2201.08825}
\BIBentrySTDinterwordspacing

\bibitem[Mitarai and Fujii(2021{\natexlab{a}})]{Mitarai_2021}
\BIBentryALTinterwordspacing
K.~Mitarai and K.~Fujii, ``Overhead for simulating a non-local channel with
  local channels by quasiprobability sampling,'' \emph{Quantum}, vol.~5, p.
  388, Jan. 2021. [Online]. Available:
  \url{http://dx.doi.org/10.22331/q-2021-01-28-388}
\BIBentrySTDinterwordspacing

\bibitem[Mitarai and Fujii(2021{\natexlab{b}})]{Mitarai_2021a}
\BIBentryALTinterwordspacing
------, ``Constructing a virtual two-qubit gate by sampling single-qubit
  operations,'' \emph{New Journal of Physics}, vol.~23, no.~2, p. 023021, feb
  2021. [Online]. Available: \url{https://dx.doi.org/10.1088/1367-2630/abd7bc}
\BIBentrySTDinterwordspacing

\bibitem[Tang et~al.(2021)Tang, Tomesh, Suchara, Larson, and
  Martonosi]{Tang_2021}
\BIBentryALTinterwordspacing
W.~Tang, T.~Tomesh, M.~Suchara, J.~Larson, and M.~Martonosi, ``Cutqc: using
  small quantum computers for large quantum circuit evaluations,'' in
  \emph{Proceedings of the 26th ACM International Conference on Architectural
  Support for Programming Languages and Operating Systems}, ser. ASPLOS
  ’21.\hskip 1em plus 0.5em minus 0.4em\relax ACM, Apr. 2021. [Online].
  Available: \url{http://dx.doi.org/10.1145/3445814.3446758}
\BIBentrySTDinterwordspacing

\bibitem[Piveteau and Sutter(2023)]{Piveteau:2023}
C.~Piveteau and D.~Sutter, ``Circuit knitting with classical communication,''
  \emph{IEEE Transactions on Information Theory}, pp. 1--1, 2023.

\bibitem[Rennela et~al.(2023)Rennela, Brand, Laarman, and Dunjko]{Rennela:2023}
\BIBentryALTinterwordspacing
M.~Rennela, S.~Brand, A.~Laarman, and V.~Dunjko, ``Hybrid divide-and-conquer
  approach for tree search algorithms,'' \emph{{Quantum}}, vol.~7, p. 959, Mar.
  2023. [Online]. Available: \url{https://doi.org/10.22331/q-2023-03-23-959}
\BIBentrySTDinterwordspacing

\bibitem[Bennett and Brassard(2014)]{Bennett2014}
C.~H. Bennett and G.~Brassard, ``Quantum cryptography: Public key distribution
  and coin tossing,'' \emph{Theoretical Computer Science}, vol. 560, pp. 7--11,
  dec 2014.

\bibitem[Djordjevic(2021)]{QKD}
I.~B. Djordjevic, ``Chapter 15 - quantum key distribution,'' in \emph{Quantum
  Information Processing, Quantum Computing, and Quantum Error Correction},
  2nd~ed., I.~B. Djordjevic, Ed.\hskip 1em plus 0.5em minus 0.4em\relax
  Academic Press, 2021, pp. 703--784.

\bibitem[Cacciapuoti et~al.(2020)Cacciapuoti, Caleffi, Tafuri, Cataliotti,
  Gherardini, and Bianchi]{QuantumInternetCaccia}
A.~S. Cacciapuoti, M.~Caleffi, F.~Tafuri, F.~S. Cataliotti, S.~Gherardini, and
  G.~Bianchi, ``Quantum internet: Networking challenges in distributed quantum
  computing,'' \emph{IEEE Network}, vol.~34, no.~1, pp. 137--143, 2020.

\bibitem[Kimble(2008)]{Kimble_2008}
\BIBentryALTinterwordspacing
H.~J. Kimble, ``The quantum internet,'' \emph{Nature}, vol. 453, no. 7198, p.
  1023–1030, Jun. 2008. [Online]. Available:
  \url{http://dx.doi.org/10.1038/nature07127}
\BIBentrySTDinterwordspacing

\bibitem[Wehner et~al.(2018)Wehner, Elkouss, and
  Hanson]{doi:10.1126/science.aam9288}
\BIBentryALTinterwordspacing
S.~Wehner, D.~Elkouss, and R.~Hanson, ``Quantum internet: A vision for the road
  ahead,'' \emph{Science}, vol. 362, no. 6412, p. eaam9288, 2018. [Online].
  Available: \url{https://www.science.org/doi/abs/10.1126/science.aam9288}
\BIBentrySTDinterwordspacing

\bibitem[Daiss et~al.(2021)Daiss, Langenfeld, Welte, Distante, Thomas, Hartung,
  Morin, and Rempe]{60m_optical_fiber}
\BIBentryALTinterwordspacing
S.~Daiss, S.~Langenfeld, S.~Welte, E.~Distante, P.~Thomas, L.~Hartung,
  O.~Morin, and G.~Rempe, ``A quantum-logic gate between distant
  quantum-network modules,'' \emph{Science}, vol. 371, no. 6529, pp. 614--617,
  2021. [Online]. Available:
  \url{https://www.science.org/doi/abs/10.1126/science.abe3150}
\BIBentrySTDinterwordspacing

\bibitem[Liu et~al.(2023)Liu, Hu, Zhu, Zhang, Xiao, Miao, Ou, Liu, Zhou, Li,
  and Guo]{liu2023distributed}
X.~Liu, X.-M. Hu, T.-X. Zhu, C.~Zhang, Y.-X. Xiao, J.-L. Miao, Z.-W. Ou, B.-H.
  Liu, Z.-Q. Zhou, C.-F. Li, and G.-C. Guo, ``Distributed quantum computing
  over 7.0 km,'' 2023.

\bibitem[Caleffi et~al.(2022)Caleffi, Amoretti, Ferrari, Cuomo, Illiano,
  Manzalini, and Cacciapuoti]{caleffi2022distributed}
M.~Caleffi, M.~Amoretti, D.~Ferrari, D.~Cuomo, J.~Illiano, A.~Manzalini, and
  A.~S. Cacciapuoti, ``{Distributed Quantum Computing: a Survey},'' 2022.

\bibitem[Barral et~al.(2024)Barral, Cardama, Díaz, Faílde, Llovo, Juane,
  Vázquez-Pérez, Villasuso, Piñeiro, Costas, Pichel, Pena, and
  Gómez]{barral2024reviewdistributedquantumcomputing}
\BIBentryALTinterwordspacing
D.~Barral, F.~J. Cardama, G.~Díaz, D.~Faílde, I.~F. Llovo, M.~M. Juane,
  J.~Vázquez-Pérez, J.~Villasuso, C.~Piñeiro, N.~Costas, J.~C. Pichel, T.~F.
  Pena, and A.~Gómez, ``Review of distributed quantum computing. from single
  qpu to high performance quantum computing,'' 2024. [Online]. Available:
  \url{https://arxiv.org/abs/2404.01265}
\BIBentrySTDinterwordspacing

\bibitem[Benatallah et~al.(2004)Benatallah, Casati, and
  Toumani]{Benatallah:2004}
B.~Benatallah, F.~Casati, and F.~Toumani, ``Web service conversation modeling:
  a cornerstone for e-business automation,'' \emph{IEEE Internet Computing},
  vol.~8, no.~1, pp. 46--54, 2004.

\bibitem[Canetti et~al.(1996)Canetti, Feige, Goldreich, and Naor]{Canetti:1996}
R.~Canetti, U.~Feige, O.~Goldreich, and M.~Naor, ``Adaptively secure
  multi-party computation,'' in \emph{Proceedings of the Twenty-Eighth Annual
  ACM Symposium on Theory of Computing}, ser. STOC '96.\hskip 1em plus 0.5em
  minus 0.4em\relax New York, NY, USA: Association for Computing Machinery,
  1996, p. 639–648.

\bibitem[Neumann and Wezeman(2022)]{Neumann2022}
N.~M.~P. Neumann and R.~S. Wezeman, ``Distributed quantum machine learning,''
  in \emph{Innovations for Community Services}.\hskip 1em plus 0.5em minus
  0.4em\relax Springer International Publishing, 2022, pp. 281--293.

\bibitem[Zhang et~al.(2022)Zhang, Li, and Yuan]{prepare_quantum_state}
\BIBentryALTinterwordspacing
X.-M. Zhang, T.~Li, and X.~Yuan, ``Quantum state preparation with optimal
  circuit depth: Implementations and applications,'' \emph{Phys. Rev. Lett.},
  vol. 129, p. 230504, Nov 2022. [Online]. Available:
  \url{https://link.aps.org/doi/10.1103/PhysRevLett.129.230504}
\BIBentrySTDinterwordspacing

\bibitem[Lloyd et~al.(2020)Lloyd, Schuld, Ijaz, Izaac, and
  Killoran]{lloyd2020quantum}
S.~Lloyd, M.~Schuld, A.~Ijaz, J.~Izaac, and N.~Killoran, ``Quantum embeddings
  for machine learning,'' 2020.

\bibitem[Sarker(2021)]{Sarker2021}
I.~H. Sarker, ``Machine learning: Algorithms, real-world applications and
  research directions,'' \emph{SN Computer Science}, vol.~2, no.~3, p. 160,
  2021.

\bibitem[OpenAI(2022{\natexlab{a}})]{dall-e2}
OpenAI, ``{DALL}-{E}2,'' URL: \url{https://openai.com/dall-e-2}, 2022.

\bibitem[OpenAI(2022{\natexlab{b}})]{chatgpt}
------, ``Chat{GPT},'' URL: \url{https://openai.com/chatgpt}, 2022.

\bibitem[Narayanan et~al.(2021)Narayanan, Shoeybi, Casper, LeGresley, Patwary,
  Korthikanti, Vainbrand, Kashinkunti, Bernauer, Catanzaro, Phanishayee, and
  Zaharia]{Narayanan2021}
D.~Narayanan, M.~Shoeybi, J.~Casper, P.~LeGresley, M.~Patwary, V.~Korthikanti,
  D.~Vainbrand, P.~Kashinkunti, J.~Bernauer, B.~Catanzaro, A.~Phanishayee, and
  M.~Zaharia, ``{Efficient Large-Scale Language Model Training on GPU Clusters
  Using Megatron-LM},'' in \emph{Proceedings of the International Conference
  for High Performance Computing, Networking, Storage and Analysis}, ser. SC
  '21.\hskip 1em plus 0.5em minus 0.4em\relax New York, NY, USA: Association
  for Computing Machinery, 2021.

\bibitem[Schuld and Petruccione(2021)]{schuld2021machine}
M.~Schuld and F.~Petruccione, \emph{Machine Learning with Quantum Computers},
  ser. Quantum Science and Technology.\hskip 1em plus 0.5em minus 0.4em\relax
  Springer Cham, 2021.

\bibitem[Farhi et~al.(2014)Farhi, Goldstone, and Gutmann]{farhi2014quantum}
E.~Farhi, J.~Goldstone, and S.~Gutmann, ``{A Quantum Approximate Optimization
  Algorithm},'' 2014.

\bibitem[de~Wolf(2023)]{dewolf2023quantum}
R.~de~Wolf, ``Quantum computing: Lecture notes,'' 2023.

\bibitem[Cerezo et~al.(2021)Cerezo, Arrasmith, Babbush, Benjamin, Endo, Fujii,
  McClean, Mitarai, Yuan, Cincio, and Coles]{Cerezo_2021}
M.~Cerezo, A.~Arrasmith, R.~Babbush, S.~C. Benjamin, S.~Endo, K.~Fujii, J.~R.
  McClean, K.~Mitarai, X.~Yuan, L.~Cincio, and P.~J. Coles, ``Variational
  quantum algorithms,'' \emph{Nature Reviews Physics}, vol.~3, no.~9, pp.
  625--644, aug 2021.

\bibitem[DiAdamo et~al.(2021)DiAdamo, Ghibaudi, and Cruise]{DiAdamo_2021}
S.~DiAdamo, M.~Ghibaudi, and J.~Cruise, ``Distributed quantum computing and
  network control for accelerated {VQE},'' \emph{{IEEE} Transactions on Quantum
  Engineering}, vol.~2, pp. 1--21, 2021.

\bibitem[Khait et~al.(2023)Khait, Tham, Segal, and
  Brodutch]{khait2023variational}
I.~Khait, E.~Tham, D.~Segal, and A.~Brodutch, ``{Variational Quantum
  Eigensolvers in the Era of Distributed Quantum Computers},'' 2023.

\bibitem[Du et~al.(2022)Du, Qian, Wu, and
  Tao]{distributed_learning_scheme_vqas}
Y.~Du, Y.~Qian, X.~Wu, and D.~Tao, ``s,'' \emph{IEEE Transactions on Quantum
  Engineering}, vol.~3, pp. 1--16, 2022.

\bibitem[Li et~al.(2023)Li, Alam, and Ghosh]{large_scale_qaoa}
J.~Li, M.~Alam, and S.~Ghosh, ``{Large-Scale Quantum Approximate Optimization
  via Divide-and-Conquer},'' \emph{IEEE Transactions on Computer-Aided Design
  of Integrated Circuits and Systems}, vol.~42, no.~6, pp. 1852--1860, 2023.

\bibitem[Niu et~al.(2023)Niu, Zhang, Ding, Bao, and Huang]{Niu2023}
Y.-F. Niu, S.~Zhang, C.~Ding, W.-S. Bao, and H.-L. Huang, ``Parameter-parallel
  distributed variational quantum algorithm,'' \emph{{SciPost} Physics},
  vol.~14, no.~5, May 2023.

\bibitem[Jozsa(2005)]{jozsa2005introduction}
R.~Jozsa, ``An introduction to measurement based quantum computation,'' 2005.

\bibitem[Grover(1996)]{Grover:1996}
L.~K. Grover, ``{A Fast Quantum Mechanical Algorithm for Database Search},'' in
  \emph{Proceedings of the Twenty-Eighth Annual ACM Symposium on Theory of
  Computing}, ser. STOC '96.\hskip 1em plus 0.5em minus 0.4em\relax New York,
  NY, USA: Association for Computing Machinery, 1996, p. 212–219.

\bibitem[Shor(1997)]{Shor_1997}
P.~W. Shor, ``{Polynomial-Time Algorithms for Prime Factorization and Discrete
  Logarithms on a Quantum Computer},'' \emph{{SIAM} J. Comput.}, vol.~26,
  no.~5, pp. 1484--1509, 1997.

\bibitem[Zhou et~al.(2023)Zhou, Qiu, and Luo]{Zhou2023}
X.~Zhou, D.~Qiu, and L.~Luo, ``{Distributed exact Grover's algorithm},''
  \emph{Frontiers of Physics}, vol.~18, no.~5, Aug. 2023.

\bibitem[Yimsiriwattana and Jr.(2004)]{Yimsiriwattana:Shor:2004}
A.~Yimsiriwattana and S.~J.~L. Jr., ``{Distributed quantum computing: a
  distributed Shor algorithm},'' in \emph{Quantum Information and Computation
  II}, E.~Donkor, A.~R. Pirich, and H.~E. Brandt, Eds., vol. 5436,
  International Society for Optics and Photonics.\hskip 1em plus 0.5em minus
  0.4em\relax SPIE, 2004, pp. 360 -- 372.

\bibitem[Hariharan(2006)]{HARIHARAN2007}
P.~Hariharan, \emph{\BIBforeignlanguage{eng}{Basics of interferometry}},
  2nd~ed.\hskip 1em plus 0.5em minus 0.4em\relax Boston: Elsevier Academic
  Press, 2006.

\bibitem[{Boffin} et~al.(2016){Boffin}, {Hussain}, {Berger}, and
  {Schmidtobreick}]{Boffin2016}
H.~M.~J. {Boffin}, G.~{Hussain}, J.-P. {Berger}, and L.~{Schmidtobreick}, Eds.,
  \emph{{Astronomy at High Angular Resolution}}, ser. Astrophysics and Space
  Science Library, vol. 439, 2016.

\bibitem[Cross et~al.(2019)Cross, Bishop, Sheldon, Nation, and
  Gambetta]{QuantumVolume:2019}
A.~W. Cross, L.~S. Bishop, S.~Sheldon, P.~D. Nation, and J.~M. Gambetta,
  ``Validating quantum computers using randomized model circuits,'' \emph{Phys.
  Rev. A}, vol. 100, p. 032328, Sep 2019.

\bibitem[Otero et~al.(2023)Otero, Neumann, van~der Schoot, and
  Wezeman]{RNSW:2023}
A.~R. Otero, N.~M.~P. Neumann, W.~van~der Schoot, and R.~Wezeman, ``Noise
  robustness of a multiparty quantum summation protocol,'' 2023, accepted at
  ICCS 2024.

\bibitem[H{\o}yer and {\v S}palek(2005)]{HS2005}
P.~H{\o}yer and R.~{\v S}palek, ``Quantum fan-out is powerful,'' \emph{Theory
  of Computing}, vol.~1, no.~5, pp. 81--103, 2005.

\bibitem[Eisert et~al.(2000)Eisert, Jacobs, Papadopoulos, and
  Plenio]{Eisert2000}
J.~Eisert, K.~Jacobs, P.~Papadopoulos, and M.~B. Plenio, ``Optimal local
  implementation of nonlocal quantum gates,'' \emph{Phys. Rev. A}, vol.~62, p.
  052317, Oct 2000.

\bibitem[Yimsiriwattana and Lomonaco(2004)]{Yimsiriwattana2004}
A.~Yimsiriwattana and S.~J. Lomonaco, ``{Generalized GHZ States and Distributed
  Quantum Computing},'' 2004.

\bibitem[Wang et~al.(2022)Wang, Li, Xue, Cheng, Yu, Sun, and
  Lu]{asynchronous_network}
Z.~Wang, J.~Li, K.~Xue, S.~Cheng, N.~Yu, Q.~Sun, and J.~Lu, ``{An Asynchronous
  Entanglement Distribution Protocol for Quantum Networks},'' \emph{IEEE
  Network}, vol.~36, no.~5, pp. 40--47, 2022.

\bibitem[Zhang and James(2011)]{Zhang_2011}
G.~Zhang and M.~R. James, ``{Direct and Indirect Couplings in Coherent Feedback
  Control of Linear Quantum Systems},'' \emph{{IEEE} Transactions on Automatic
  Control}, vol.~56, no.~7, pp. 1535--1550, jul 2011.

\bibitem[Xiang et~al.(2013)Xiang, Ashhab, You, and Nori]{Xiang_2013}
Z.-L. Xiang, S.~Ashhab, J.~Q. You, and F.~Nori, ``Hybrid quantum circuits:
  Superconducting circuits interacting with other quantum systems,''
  \emph{Reviews of Modern Physics}, vol.~85, no.~2, pp. 623--653, apr 2013.

\bibitem[Hafezi et~al.(2009)Hafezi, Chang, Gritsev, Demler, and
  Lukin]{Hafezi_optical}
M.~Hafezi, D.~Chang, V.~Gritsev, E.~Demler, and M.~Lukin, ``Photonic quantum
  transport in a nonlinear optical fiber,'' \emph{Europhysics Letters (epl)},
  vol.~94, 07 2009.

\bibitem[Herbst et~al.(2015)Herbst, Scheidl, Fink, Handsteiner, Wittmann,
  Ursin, and Zeilinger]{Herbst_2015}
T.~Herbst, T.~Scheidl, M.~Fink, J.~Handsteiner, B.~Wittmann, R.~Ursin, and
  A.~Zeilinger, ``Teleportation of entanglement over 143 km,''
  \emph{Proceedings of the National Academy of Sciences}, vol. 112, no.~46, pp.
  14\,202--14\,205, nov 2015.

\bibitem[Chou et~al.(2018)Chou, Blumoff, Wang, Reinhold, Axline, Gao, Frunzio,
  Devoret, Jiang, and Schoelkopf]{Chou_2018}
\BIBentryALTinterwordspacing
K.~S. Chou, J.~Z. Blumoff, C.~S. Wang, P.~C. Reinhold, C.~J. Axline, Y.~Y. Gao,
  L.~Frunzio, M.~H. Devoret, L.~Jiang, and R.~J. Schoelkopf, ``Deterministic
  teleportation of a quantum gate between two logical qubits,'' \emph{Nature},
  vol. 561, no. 7723, p. 368–373, Sep. 2018. [Online]. Available:
  \url{http://dx.doi.org/10.1038/s41586-018-0470-y}
\BIBentrySTDinterwordspacing

\bibitem[Zhong et~al.(2021)Zhong, Chang, Bienfait, Dumur, Chou, Conner, Grebel,
  Povey, Yan, Schuster, and Cleland]{Zhong_2021}
Y.~Zhong, H.-S. Chang, A.~Bienfait, {\'{E}}.~Dumur, M.-H. Chou, C.~R. Conner,
  J.~Grebel, R.~G. Povey, H.~Yan, D.~I. Schuster, and A.~N. Cleland,
  ``Deterministic multi-qubit entanglement in a quantum network,''
  \emph{Nature}, vol. 590, no. 7847, pp. 571--575, feb 2021.

\bibitem[Noel et~al.(2022)Noel, Niroula, Zhu, Risinger, Egan, Biswas, Cetina,
  Gorshkov, Gullans, Huse, and Monroe]{Noel_2022}
C.~Noel, P.~Niroula, D.~Zhu, A.~Risinger, L.~Egan, D.~Biswas, M.~Cetina, A.~V.
  Gorshkov, M.~J. Gullans, D.~A. Huse, and C.~Monroe, ``Measurement-induced
  quantum phases realized in a trapped-ion quantum computer,'' \emph{Nature
  Physics}, vol.~18, no.~7, pp. 760--764, jun 2022.

\bibitem[Saffman(2016)]{Saffman_2016}
M.~Saffman, ``Quantum computing with atomic qubits and {Rydberg} interactions:
  progress and challenges,'' \emph{Journal of Physics B: Atomic, Molecular and
  Optical Physics}, vol.~49, no.~20, p. 202001, oct 2016.

\bibitem[Struck and Burkard(2016)]{Struck2016}
P.~R. Struck and G.~Burkard, \emph{Spin Quantum Computing}.\hskip 1em plus
  0.5em minus 0.4em\relax Dordrecht: Springer Netherlands, 2016, pp. 71--103.

\bibitem[Kjaergaard et~al.(2020)Kjaergaard, Schwartz, Braum{\"{u}}ller, Krantz,
  Wang, Gustavsson, and Oliver]{annurev_soa}
M.~Kjaergaard, M.~E. Schwartz, J.~Braum{\"{u}}ller, P.~Krantz, J.~I.-J. Wang,
  S.~Gustavsson, and W.~D. Oliver, ``{Superconducting Qubits: Current State of
  Play},'' \emph{Annual Review of Condensed Matter Physics}, vol.~11, no.~1,
  pp. 369--395, 2020.

\bibitem[Orlando et~al.(1999)Orlando, Mooij, Tian, van~der Wal, Levitov, Lloyd,
  and Mazo]{orlando_flux}
T.~P. Orlando, J.~E. Mooij, L.~Tian, C.~H. van~der Wal, L.~S. Levitov,
  S.~Lloyd, and J.~J. Mazo, ``Superconducting persistent-current qubit,''
  \emph{Phys. Rev. B}, vol.~60, pp. 15\,398--15\,413, Dec 1999.

\bibitem[Lachance-Quirion et~al.(2019)Lachance-Quirion, Tabuchi, Gloppe, Usami,
  and Nakamura]{Lachance_Quirion_2019}
D.~Lachance-Quirion, Y.~Tabuchi, A.~Gloppe, K.~Usami, and Y.~Nakamura, ``Hybrid
  quantum systems based on magnonics,'' \emph{Applied Physics Express},
  vol.~12, no.~7, p. 070101, jun 2019.

\bibitem[Psaroudaki and Panagopoulos(2021)]{Psaroudaki_2021}
C.~Psaroudaki and C.~Panagopoulos, ``{Skyrmion Qubits: A New Class of Quantum
  Logic Elements Based on Nanoscale Magnetization},'' \emph{Physical Review
  Letters}, vol. 127, no.~6, aug 2021.

\bibitem[DiVincenzo(2000)]{DiVincenzo_2000}
D.~P. DiVincenzo, ``{The Physical Implementation of Quantum Computation},''
  \emph{Fortschritte der Physik}, vol.~48, no. 9-11, pp. 771--783, sep 2000.

\bibitem[Lago-Rivera et~al.(2023)Lago-Rivera, Rakonjac, Grandi, and
  de~Riedmatten]{Lago_Rivera_2023}
D.~Lago-Rivera, J.~V. Rakonjac, S.~Grandi, and H.~de~Riedmatten, ``Long
  distance multiplexed quantum teleportation from a telecom photon to a
  solid-state qubit,'' \emph{Nature Communications}, vol.~14, no.~1, apr 2023.

\bibitem[Palani et~al.(2023)Palani, Hasse, Kiefer, Boeckling, Schroeder,
  Warring, and Schaetz]{Palani_2023}
D.~Palani, F.~Hasse, P.~Kiefer, F.~Boeckling, J.-P. Schroeder, U.~Warring, and
  T.~Schaetz, ``High-fidelity transport of trapped-ion qubits in a multilayer
  array,'' \emph{Physical Review A}, vol. 107, no.~5, may 2023.

\bibitem[van~der Schoot et~al.(2023)van~der Schoot, Wezeman, Eendebak, Neumann,
  and Phillipson]{levels_quantum_metrics:2023}
\BIBentryALTinterwordspacing
W.~van~der Schoot, R.~Wezeman, P.~T. Eendebak, N.~M.~P. Neumann, and
  F.~Phillipson, ``Evaluating three levels of quantum metrics on
  quantum-inspire hardware,'' \emph{Quantum Information Processing}, vol.~22,
  no.~12, Dec. 2023. [Online]. Available:
  \url{http://dx.doi.org/10.1007/s11128-023-04184-x}
\BIBentrySTDinterwordspacing

\end{thebibliography}

% Use below for regular version (no \citeauthor and similar)
% \bibliographystyle{IEEEtran} %
% \bibliography{references}

\end{document}